\documentstyle[11pt,newpasp,twoside,epsf]{article}
\def\edcomment#1{\iffalse\marginpar{\raggedright\sl#1\/}\else\relax\fi}
\reversemarginpar

\begin{document}
\title{An Adaptive Optics Survey for Companions to Stars with Extra-Solar Planets}
\author{James~P.~Lloyd, Michael~C.~Liu, James~R.~Graham, Melissa~Enoch, Paul~Kalas, Geoffrey~W.~Marcy, Debra~Fischer}
\affil{Department of Astronomy, University of California, Berkeley, CA 94720, USA}
\author{Jennifer~Patience, Bruce~Macintosh, Donald~T.~Gavel, Scot~S.~Olivier, Claire~E.~Max}
\affil{Institute of Geophysics and Planetary Physics, Lawrence Livermore National Laboratory, Livermore, CA 95064, USA}
\author{Russel~White, Andrea~M.~Ghez, Ian~S.~McLean}
\affil{Department of Physics and Astronomy, University of California, Los Angeles, CA 90095, USA}

\begin{abstract}
We have undertaken an adaptive optics imaging survey of extrasolar planetary systems and stars 
showing interesting radial velocity trends from high precision radial velocity 
searches.  
Adaptive Optics increases the resolution 
and dynamic range of an image, substantially improving the detectability 
of faint close companions.  This survey is sensitive to objects less luminous than
the bottom of the main sequence at separations as close as 1\arcsec.  
We have detected stellar companions to the planet 
bearing stars HD 114762 and Tau Boo.  We have also detected a companion 
to the non-planet bearing star 16 Cyg A.
\end{abstract}

\section{Introduction}

The Lick Adaptive Optics System is a Shack-Hartmann Laser Guide star AO
system on the Lick Observatory Shane 3m telescope at 
Mt Hamilton, California (Max et al. 1997).  
For these observations, we used the system 
in natural guide star mode, with the bright star serving as wavefront
reference.  Under good conditions, the system produces diffraction
limited images (0\farcs 15 FWHM) with a strehl ratio of 0.7 in the K band (2.2 $\mu$m).

The AO system feeds a 256$\times$256 pixel infrared camera, IRCAL (Lloyd et al. 2000), 
which reimages the field of view at 0\farcs 076 per pixel.  
The camera incorporates a cold focal plane 
with an occulting finger to obtain high dynamic range images.  For
this program, we typically take a few minutes of integration of 
unsaturated images to obtain coverage close to the star, and 
deep exposures in coronagraphic mode to detect faint companions at
larger separations.

We have selected targets from those stars with radial velocity 
planets, or 
with radial velocity trends from the Lick and Keck 
radial 
velocity surveys.  

\section{Results} 

HD 155423 (see Fig 1) shows substantial scatter in
precision radial velocity measurements.  High resolution imaging shows HD 155423
is a hierarchical triple, with a 0\farcs 2 close binary (8AU projected separation) M3 dwarf pair, separated 1\farcs 5 from the F8 dwarf primary.

A faint object was previously discovered near 16 Cyg A, but
was not known to be physically associated (Hauser \& Marcy 1999).  We
have confirmed by proper motion measurements that the $\Delta$K=5.4 object
is a physically associated M5 dwarf (see Fig 1).  
Radial velocity measurements show a shallow linear trend.

HD 114762 has a radial velocity companion with an 84
day period and M$\sin i$=11 M$_{Jup}$ (Latham et al. 1989).  We have detected
a $\Delta$K=7.3 companion 3\farcs 3 from the primary (see Fig 1).
{\it JHK} photometry and follow up Keck AO/NIRSPEC spectroscopy (McLean et al. 2000) reveal the companion to be a 
late M subdwarf.  The physical association of this companion is confirmed by
proper motion over a 2 year baseline (Patience et al. 1998).

Radial velocity measurements of HR 6623 show a nearly linear trend over
13 years (see Fig 1).  This would classify this object
as a poorly determined single lined spectroscopic binary.  AO imaging
resolves the companion, which is an M5 dwarf.  Further radial velocity
and astrometric observations will allow accurate mass determinations.

Tau Boo hosts an M$\sin i$ = 3.9 M$_{Jup}$, 3.3 day period planet, and shows radial velocity residuals (see Fig 1).  It has
an M2V companion that was discovered in 1849 at 10\farcs 3.  At 
present it is at 2\farcs 82, with 0\farcs 01 per year of orbital motion.  
Although it has been suggested that there may be additional companions
in the system (Wiedemann, Deming, \& Bjoraker 2000), we do not
detect any additional companions, and attribute the velocity residuals to the 
stellar companion.

\begin{figure} 
\plotone{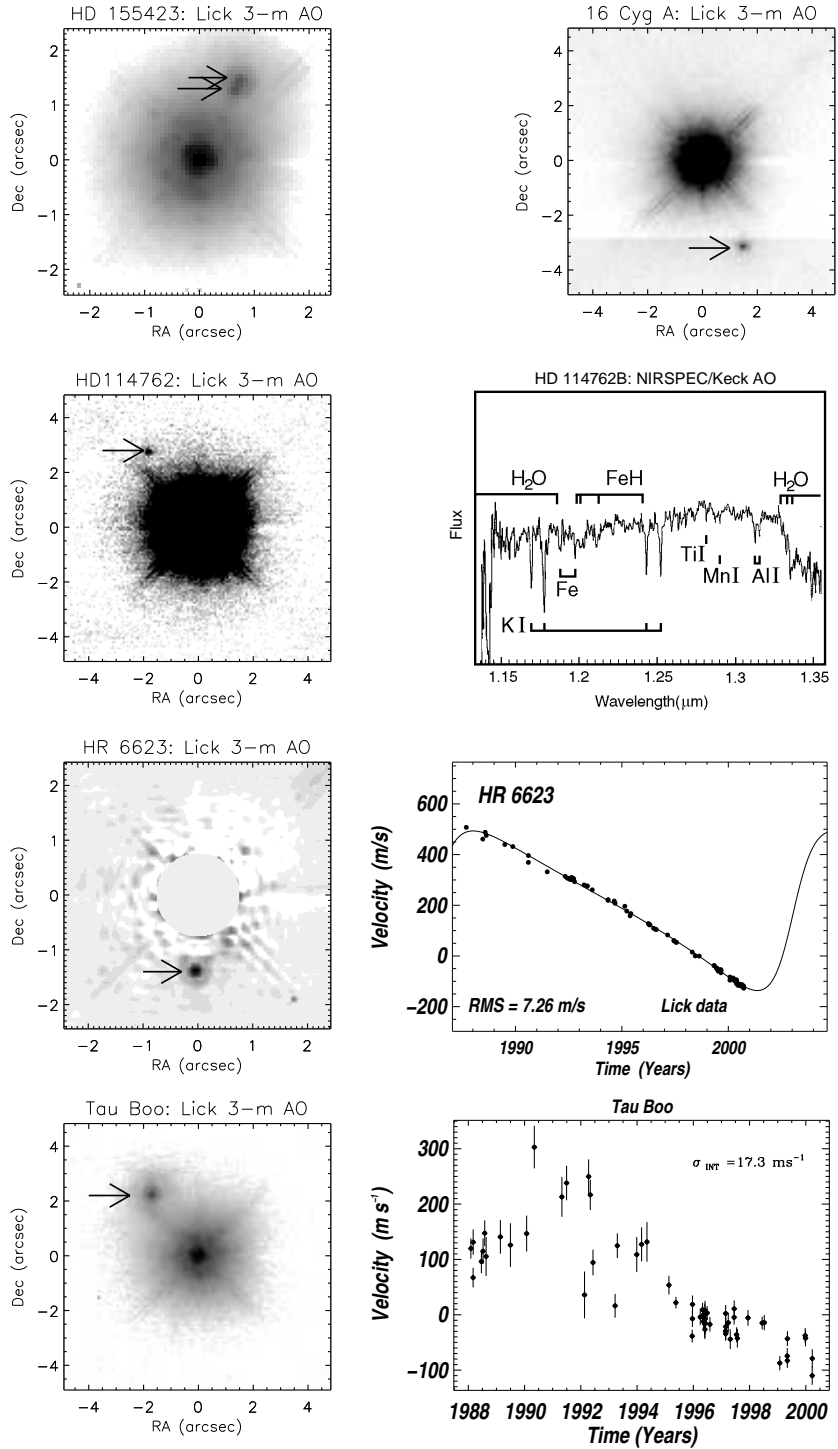}
\caption{K band images of stars with detected companions; J band spectra of HD 114762B; Radial Velocity data for HR 6623 and Tau Boo (residual)  }
\end{figure}

\acknowledgements

This research was supported in part by the National 
Science Foundation under a cooperative agreement with the Center for 
Adaptive Optics, Agreement No. AST-987678.   
Work on the Lick adaptive optics system was performed in
part under the auspices of the U.S. Department of Energy by
Lawrence Livermore National Laboratory, Contract
number W-7405-ENG-48
 
\end{document}